%% file: NonGaussianEntangledState-arxiv.tex
\documentclass[twocolumn,aps,prl,noshowpacs,amsmath,amssymb,superscriptaddress]{revtex4}
\usepackage{graphicx}
\usepackage{amssymb}
\usepackage{amsmath}
\usepackage{float}
\usepackage{bibunits}
\usepackage{textcomp}

\newcommand{\ket}[1]{\ensuremath{|{#1}\rangle}}
\newcommand{\bracket}[2]{\ensuremath{\langle{#1}|{#2}\rangle}}

\newcommand{\aver}[1]{\ensuremath{\langle  #1 \rangle}}
\newcommand{\var}[1]{\ensuremath{\langle  #1^2 \rangle}}

\newcommand{\hh}[0]{\ensuremath{\ket{\textrm{h}}}}
\newcommand{\vv}[0]{\ensuremath{\ket{\textrm{v}}}}

\def\ba{\begin{eqnarray}}
\def\ea{\end{eqnarray}}
\def\beq{\begin{equation}}
\def\eeq{\end{equation}}
\def\beqstar{\begin{equation*}}
\def\eeqstar{\end{equation*}}

\begin{document}

\begin{bibunit}

\title{Entanglement with Negative Wigner Function of Three \\Thousand Atoms Heralded by One Photon}

\author{Robert McConnell}
\affiliation{
Department of Physics, MIT-Harvard Center for Ultracold Atoms,
and Research Laboratory of Electronics, Massachusetts Institute of Technology,
Cambridge, Massachusetts 02139, USA}

\author{Hao Zhang}
\affiliation{
Department of Physics, MIT-Harvard Center for Ultracold Atoms,
and Research Laboratory of Electronics, Massachusetts Institute of Technology,
Cambridge, Massachusetts 02139, USA}

\author{Jiazhong Hu}
\affiliation{
Department of Physics, MIT-Harvard Center for Ultracold Atoms,
and Research Laboratory of Electronics, Massachusetts Institute of Technology,
Cambridge, Massachusetts 02139, USA}

\author{Senka \'{C}uk}
\affiliation{
Department of Physics, MIT-Harvard Center for Ultracold Atoms,
and Research Laboratory of Electronics, Massachusetts Institute of Technology,
Cambridge, Massachusetts 02139, USA}
\affiliation{
Institute of Physics, University of Belgrade, Pregrevica 118, 11080 Belgrade, Serbia}

\author{Vladan Vuleti\'{c}}
\affiliation{
Department of Physics, MIT-Harvard Center for Ultracold Atoms,
and Research Laboratory of Electronics, Massachusetts Institute of Technology,
Cambridge, Massachusetts 02139, USA}

\date{\today}

\begin{abstract}
Quantum-mechanically correlated (entangled) states of many particles are of interest in quantum information, quantum computing and quantum metrology. Metrologically useful entangled states of large atomic ensembles have been experimentally realized\cite{Kitagawa93, Appel09,Takano10,Schleier-Smith10, Leroux2010, Gross2010, Riedel2010,Hamley2012,Sewell12,Bohnet14}, but these states display Gaussian spin distribution functions with a non-negative Wigner function. Non-Gaussian entangled states have been produced in small ensembles of ions\cite{Leibfried2006,Monz11}, and very recently in large atomic ensembles\cite{Haas2014,Strobel2014,Luecke2014}. Here, we generate entanglement in a large atomic ensemble via the interaction with a very weak laser pulse; remarkably, the detection of a single photon prepares several thousand atoms in an entangled state. We reconstruct a negative-valued Wigner function, an important hallmark of nonclassicality, and verify an entanglement depth (minimum number of mutually entangled atoms) of $2910 \pm 190$ out of $3100$ atoms. This is the first time a negative Wigner function or the mutual entanglement of virtually all atoms have been attained in an ensemble containing more than a few particles. While the achieved purity of the state is slightly below the threshold for entanglement-induced metrological gain, further technical improvement should allow the generation of states that surpass this threshold, and of more complex Schr\"odinger cat states for quantum metrology and information processing. More generally, our results demonstrate the power of heralded methods for entanglement generation, and illustrate how the information contained in a single photon can drastically alter the quantum state of a large system.
\end{abstract}

\maketitle

Entanglement is now recognized as a resource for secure communication, quantum information processing, and precision measurements. An important goal is the creation of entangled states of many-particle systems while retaining the ability to characterize the quantum state and validate entanglement. Entanglement can be verified in a variety of ways, with one of the strictest criteria being a negative-valued Wigner function\cite{Leibfried96,Lvovsky2001}, that necessarily implies that the entangled state has a non-Gaussian wavefunction. To date, the metrologically useful spin-squeezed states\cite{Kitagawa93, Appel09,Takano10,Schleier-Smith10, Leroux2010, Gross2010, Riedel2010,Hamley2012,Sewell12,Bohnet14} have been produced in large ensembles. These states have Gaussian spin distributions and therefore can largely be modeled as systems with a classical source of spin noise, where quantum mechanics enters only to set the amount of Gaussian noise. Non-Gaussian states with a negative Wigner function, however, are manifestly non-classical, since the Wigner function as a quasiprobability function must remain non-negative in the classical realm. While prior to this work a negative Wigner function had not been attained for atomic ensembles, in the optical domain, a negative-valued Wigner function has very recently been measured for states with up to 110 microwave photons\cite{Vlastakis2013}. Another entanglement measure is the entanglement depth\cite{Sorensen01}, i.e. the minimum number of atoms that are demonstrably, but possibly weakly, entangled with one another. This parameter quantifies how widely shared among the particles an entangled state is. For a state of an ensemble characterized by collective measurements, the entanglement depth depends sensitively on the proximity of the state to the ideal symmetric subspace of all particles. The largest entanglement depth verified previously has been 170 out of 2300 atoms for a spin-squeezed state\cite{Gross2010}, and very recently 13 out of 41 atoms for a non-Gaussian state\cite{Haas2014}.

Here we generate entanglement in a large atomic ensemble by detecting a single photon that has interacted with the ensemble\cite{McConnell2013}. An incident vertically polarized photon experiences a weak random polarization rotation associated with the quantum noise of the collective atomic spin. The detection of a horizontally polarized emerging photon then heralds a non-Gaussian entangled state of collective atomic spin (Fig. 1) with a negative-valued Wigner function of $-0.36 \pm 0.08$, and an entanglement depth of $90 \%$ of our ensemble containing several thousand atoms.

\begin{figure*}[ht]
\centering
\includegraphics[width=.75 \textwidth]{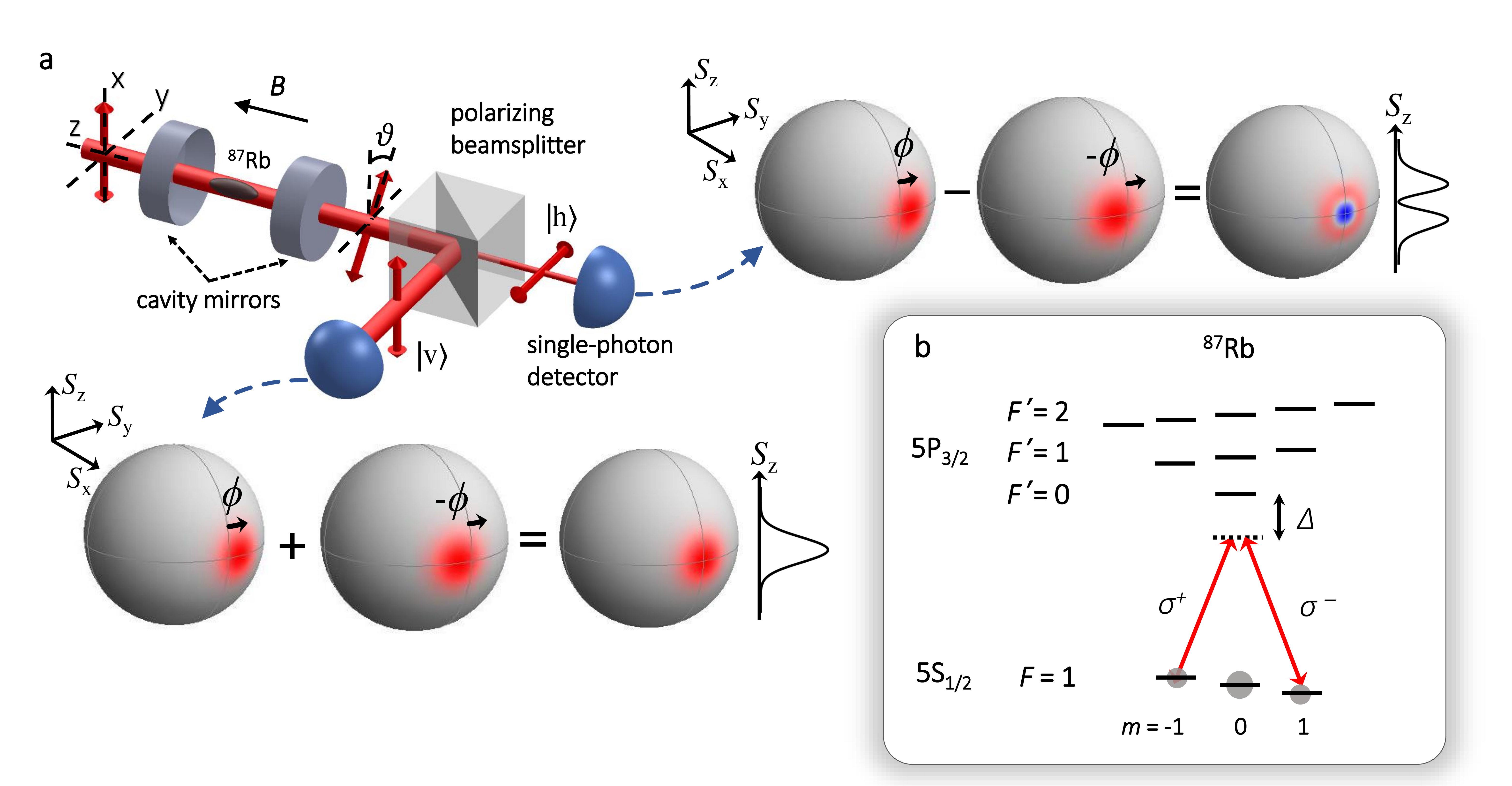}
\caption{Scheme for heralded entanglement generation in a large atomic ensemble by single-photon detection. (a) Incident vertically polarized light experiences weak polarization rotation due to atomic quantum noise, and the detection of a horizontally polarized transmitted photon heralds an entangled state of collective atomic spin. An optical resonator enhances the polarization rotation and the heralding probability. (b) Atoms in the $5 S_{1/2}, F=1$ hyperfine manifold are coupled to the excited $5 P_{3/2}$ manifold via linearly polarized light, decomposed into two circular polarization components $\ket{\sigma^{\pm}}$ that interact with the atomic ground-state populations. The outgoing polarization state of the light reflects the quantum fluctuations between the $\ket{5S_{1/2}F=1,m= \pm 1}$ magnetic sublevels.}
\label{fig:fig1}
\end{figure*}

The pertinent atom-light interaction is enhanced by an optical cavity, into which we load $N_a = 3100 \pm 300$ laser-cooled $^{87}$Rb atoms (Fig. 1a). The atoms are prepared in the $5S_{1/2}, F=1$ hyperfine manifold, such that each atom $i$ can be associated with a spin $\vec{f_i}$, and the ensemble with a collective-spin vector $\vec{S} = \sum_i \vec{f}_i$. After polarizing the ensemble ($S_z \approx S$) by optical pumping, the collective spin state is rotated onto the $\hat{x}$ axis by means of a radiofrequency $\pi/2$ pulse. This (unentangled) initial state that is centered about $S_z = 0$ with a variance $(\Delta S_z)^2 = S/2$ is known as a coherent spin state (CSS). In our experiment, the atoms are non-uniformly coupled to the optical mode used for state preparation and detection, but the relevant concepts can be generalized to this situation, as discussed in Methods.

Probe light resonant with a cavity mode and detuned from the $^{87}$Rb D$_2$ transition is polarization analyzed upon transmission through the cavity. The vertical polarization state of each photon in the incident laser pulse $\vv =(\ket{\sigma^+} + \ket{\sigma^-}) / \sqrt{2}$ can be decomposed into two circular polarization components $\ket{\sigma^\pm}$ that produce opposite differential light shifts between the atomic magnetic sublevels $\ket{m=\pm 1}$. Hence a $\ket{\sigma^\pm}$ photon causes a precession of the collective spin vector $\vec{S}$ in the $xy$ plane by a small angle $\pm \phi$ (see Methods), and we denote the corresponding slightly displaced CSS by $\ket{\pm\phi}$. Then the combined state of the atom-light system after the passage of one photon can be written as\cite{McConnell2013}
\begin{equation}
\ket{\psi} \propto \ket{\sigma^+} \ket{+\phi} + \ket{\sigma^-} \ket{-\phi}.
\label{eqn:Hamil}
\end{equation}
Conversely, atoms in the states $\ket{m=\pm 1}$ cause different phase shifts on the $\sigma^\pm$ photons, resulting in a net rotation of the photon linear polarization if the states $\ket{m=\pm 1}$ are not equally populated. Then the atomic quantum fluctuations between $\ket{m=\pm 1}$ in the CSS randomly rotate the polarization of the input photons $\vv$, giving rise to a nonzero probability $ \propto \phi^2$ for an incident $\vv$ photon to emerge in the polarization $\hh =(\ket{\sigma^+} - \ket{\sigma^-}) / \sqrt{2} $, orthogonal to its input polarization. The detection of such a ``heralding" photon projects the atomic state onto $\bracket{\textrm{h}}{\psi} \propto \ket{\phi}-\ket{-\phi}$, which is not a CSS, but an entangled state of collective spin, namely, the first excited Dicke state\cite{Arecchi1972} $\ket{\psi_1}$ along $\hat{x}$ (Fig. 1a). In contrast, if the photon is detected in its original polarization $\vv$, the atomic state is projected onto $\bracket{\textrm{v}}{\psi} \propto \ket{\phi}+\ket{-\phi}$, a state slightly spin squeezed\cite{Kitagawa93} and essentially identical to the input CSS. Thus the entangled atomic state $\ket{\psi_1}$ is post-selected by the detection of the heralding photon $\hh$.

From a different perspective, the entangled state is generated by a single-photon measurement event. The incident photon undergoes Faraday rotation by an angle $\vartheta$ proportional to the collective spin along the cavity axis, $S_z$, that exhibits quantum fluctuations around $\langle S_z \rangle = 0$. Since detection of the outgoing photon in $\hh$ is only possible if $S_z \neq 0$, such detection excludes values of $S_z$ near 0 from the spin distribution\cite{McConnell2013}, and biases the collective spin towards larger values of $|S_z|$. This creates a ``hole" in the atomic distribution near $S_z = 0$, as seen in Fig. 1a.

The mean photon number in the incident laser pulse $k \sim 210$ is chosen such that the probability for one photon to emerge in heralding polarization $\hh$ is $p \approx 0.05 \ll 1$. This ensures a very small probability $\propto p^2$ for producing a different entangled state $\ket{\psi_2}$ heralded by two photons\cite{McConnell2013}, a state which, due to our photon detection efficiency of $q = 0.3 <1$, we would (mostly) mistake for $\ket{\psi_1}$. This admixture of $\ket{\psi_2}$ to the heralded state is suppressed by a factor of $3p(1-q)\approx0.1$. Further state imperfection arises from false heralding events due to residual polarization impurity of the probe beam (independent of the atoms) of $\sim 3 \times 10^{-5} =0.1 p/k$, adding an admixture of about 10\% of the CSS to the heralded state.

\begin{figure}[H]
\centering
\includegraphics[width=.5 \textwidth]{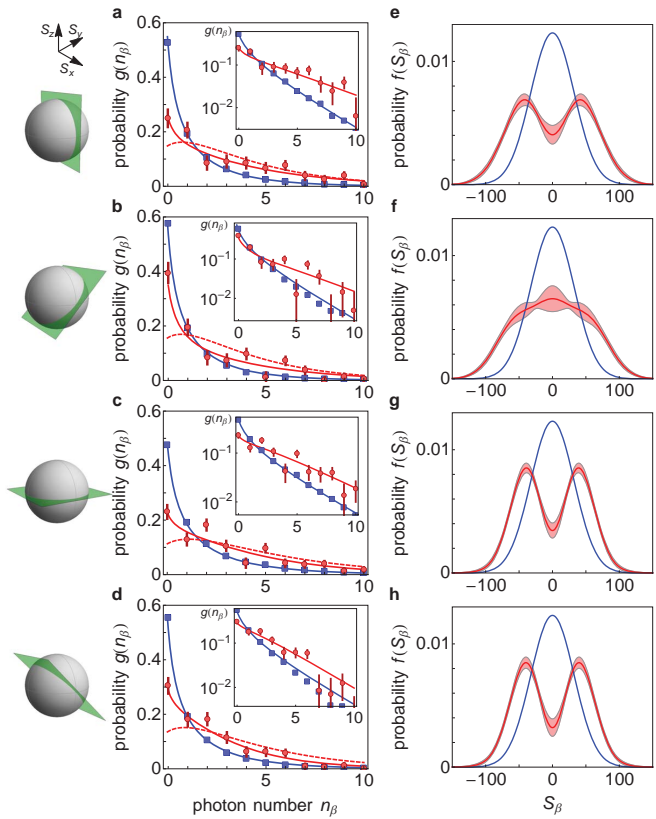}
\caption{Collective-spin distribution of atomic state heralded by one photon. (a-b) Measured photon distributions $g(n_\beta)$ for no heralding photon detected (blue squares), and for one heralding photon detected (red circles), for rotation angles (a) $\beta = 0$, (b) $\beta = \pi/4$, (c) $\beta = \pi/2$, (d) $\beta = 3\pi/4$. Inset: Logarithmic representations of the same data. In the ideal case, the ratio for the heralded state and the CSS is $\aver{n_\beta}_{\mathrm{her}}/ \aver{n_\beta}_{\mathrm{CSS}}=\aver{S_\beta^2}_{\mathrm{her}}/ \aver{S_\beta^2}_{\mathrm{CSS}}=3$ for any angle $\beta$, and we measure $\aver{n_\beta}_{\mathrm{her}}/ \aver{n_\beta}_{\mathrm{CSS}}= \{2.7 \pm 0.2, 2.2 \pm 0.2, 2.4 \pm 0.2, 2.1 \pm 0.1 \}$ for $\beta=\{ 0, \frac{\pi}{4}, \frac{\pi}{2}, \frac{3 \pi}{4} \}$. For each $\beta$, the blue and red data sets represent approximately $1.5\times10^4$ and 200 experiments, respectively. The solid blue and the dashed red curves are predictions without any free parameters, calculated from first principles and the separately measured atom number, for the CSS and the perfect first Dicke state, respectively. The solid red line corresponds to the simultaneous fit to all measurement angles $\beta$, i.e. the reconstructed density matrix. Error bars indicate 1 standard deviation (s.d.) (e-h) Reconstructed collective spin distributions of the heralded state (red) for rotation angles (e) $\beta = 0$, (f) $\beta = \pi/4$, (g) $\beta = \pi/2$, (h) $\beta = 3\pi/4$. The spin distributions of the CSS (blue) are for reference. The horizontal axis $S_z$ is expressed in terms of the effective atom number\cite{Schleier-Smith10} $N= (2/3) N_a=2100$, obtained by weighting each atom with its coupling strength to the standing-wave probe field inside the cavity, such that the experimentally measured spin fluctuation $(\Delta S_z)^2$ of the CSS via its interaction with the probe light satisfies the standard relation $(\Delta S_z)^2 = S/2 = NF/2$ for spin $F$ atoms (see Methods). The shaded area indicates the statistical uncertainty of 1 s.d.  The spin distribution in Fig. 2f shows no ``hole'' in the middle due to lower quality of data for this measurement run $\beta = \pi/4$.}
\label{fig:fig2}
\end{figure}

In order to reconstruct the collective-spin state generated by the heralding event, we rotate the atomic state after the heralding process by an angle $\beta=0, \frac{\pi}{4},\frac{\pi}{2}, \frac{3\pi}{4}$ about the $\hat{x}$ axis before measuring $S_z$. (Thus $\beta=0$ corresponds to measuring $S_z$, $\beta=\pi/2$ corresponds to $S_y$, etc.) The measurement is performed by applying a stronger light pulse in the same polarization-optimized setup used for heralding. As the Faraday rotation angle $\vartheta\ll 1$ is proportional to $S_z$, and the probability for detecting $\hh$ photons is proportional to $ \vartheta^2$, the measured probability distribution of $\hh$ photon number, $g(n_{\beta})$, reflects the probability distribution of $S_{\beta}^2$. Fig. 2a-d show that a single heralding photon substantially changes the spin distribution towards larger values of $\var{S_\beta}$. We further verify that the heralded state remains (nearly) spin polarized with a contrast of $\mathcal{C}=0.99^{+0.01}_{-0.02}$, the same as for the CSS within error bars (Fig. 3a).

From the photon distributions $g(n_\beta)$ we can reconstruct the density matrix $\rho_{mn}$ in the Dicke state basis\cite{Arecchi1972} along $\hat{x}$, where $\ket{n=0}$ denotes the CSS along $\hat{x}$, $\ket{n=1}$ the first Dicke state, $\ket{n=2}$ the second Dicke state, etc. From the density matrix we obtain the Wigner function $W(\theta,\phi)$ on the Bloch sphere\cite{Dowling1994} (Fig. \ref{fig:dmatrix}). To accurately determine the Wigner function value on the axis, $W(\theta=\frac{\pi}{2},\phi=0)= \sum_{n} (-1)^{n} \rho_{nn}$, that depends only on the population terms $\rho_{nn}$, we average the photon distributions $g(n_\beta)$ over four angles $\beta$ and thereby reduce the fitting parameters to just $\rho_{nn}, n \leq 4$. This is equivalent to constructing a rotationally symmetric Wigner function from the angle-averaged marginal distribution\cite{Lvovsky2001}. We obtain $\rho_{00}=0.32\pm0.03, \rho_{11}=0.66\pm0.04$ with negligible higher-order population terms, giving $W(\frac{\pi}{2},0) =-0.36 \pm 0.08$, to be compared to $W(\frac{\pi}{2},0)=-1$ for the perfect first Dicke state.

\begin{figure*}[t]
\centering
\includegraphics[width=.75 \textwidth]{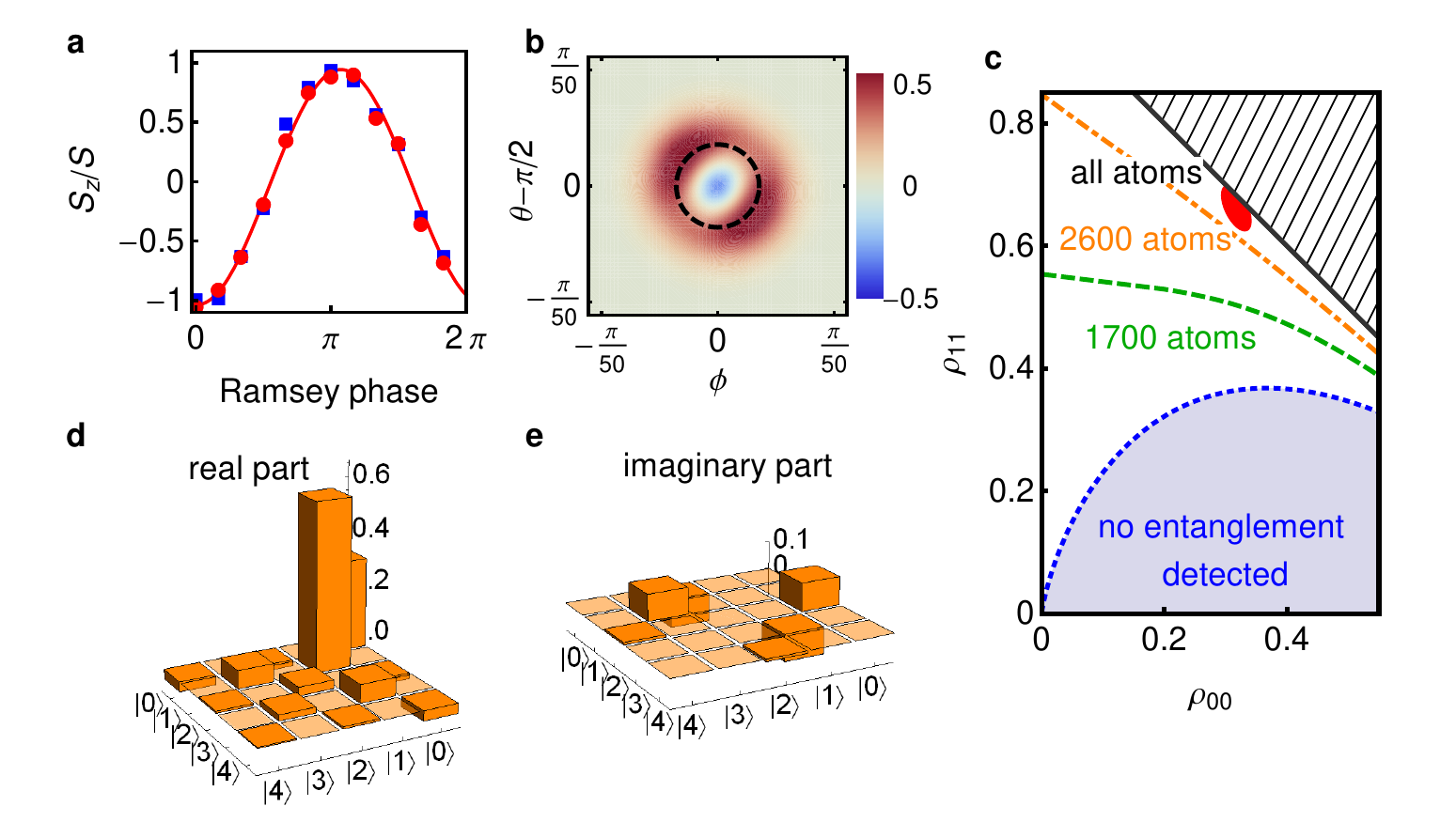}
\caption{Reconstruction of the heralded many-atom entangled state. (a) Normalized spin component $S_z/S$ measured in a Ramsey sequence, as a function of the phase of the second Ramsey $\pi/2$ pulse, for the CSS (blue squares) and the heralded state (red circles). The fit (red line) shows a contrast of $0.99^{+0.01}_{-0.0\vec{}2}$ for the heralded state, within error bars the same as the contrast $0.995\pm0.004$ of the CSS. The negligible contrast reduction is expected given that we send only 210 photons into the system at large detuning from atomic resonance. (b) Reconstructed Wigner function $W(\theta, \phi)$ for the heralded state on the Bloch sphere\cite{Dowling1994} with a radius given by the effective atom number $N=2100$. $\theta$ is the polar angle with respect to $\hat{z}$ and $\phi$ is the azimuthal angle with respect to $\hat{x}$. The first excited Dicke state and the CSS have $W(\frac{\pi}{2},0) = -1$ and $W(\frac{\pi}{2},0) = 1$, respectively. To provide a reference scale for the size of the negative region, the black dashed line is the contour at which the CSS has a Wigner function value $1/e$. (c),(d) Real and imaginary parts of the reconstructed density matrix elements, in the Dicke state basis along $\hat{x}$, for the heralded state. (e) Entanglement depth criterion\cite{Haas2014} for the heralded state, plotted in terms of density matrix elements $\rho_{00}$ and $\rho_{11}$. The red shaded region represents the $1$ s.d. confidence region for the heralded state. Lines represent boundaries for $k$-particle entanglement in terms of atom number $N_a$; a state with $\rho_{11}$ greater than such a boundary displays at least $k$-particle entanglement. States falling within the blue shaded region are not provably entangled by the used criterion. The hatched area indicates the unphysical region where the density matrix trace would exceed unity.}
\label{fig:fig3}
\end{figure*}

We can also fit the density matrix including the coherence terms simultaneously to $g(n_\beta)$ for all four angles $\beta$, without angle-averaging. Since the photon distributions $g(n_\beta)$ depend only on $S_\beta^2$, they determine only the even terms of the density matrix, i.e., $\rho_{mn}$ where $m+n$ is even, and contain no information about the odd terms. If we calculate $W(\frac{\pi}{2},0)$ from the density matrix without angle-averaging, we find $W(\frac{\pi}{2},0)=- 0.27\pm 0.08$, within error bars consistent with the angle-averaged value. In order to display the Wigner function, we bound the odd terms ($m+n$ odd) by verifying that the heralding process does not displace the state relative to the CSS (see Methods). Therefore we set the odd terms to zero, and display the resulting density matrix and corresponding Wigner function in Fig. 3b-d. The spin distributions $f(S_\beta)$ obtained from this density matrix are shown in Fig. 2e-h.

In order to quantify the minimum number of mutually entangled atoms, we use a criterion derived in Ref.\cite{Haas2014} that establishes entanglement depth as a function of the populations $\rho_{00}$ and $\rho_{11}$. From this criterion, generalized to the case of non-uniform coupling to the measurement light field (see Methods), we deduce an average entanglement depth of $\mathcal{N}_a=2910 \pm190$ out of $N_a=3100$ atoms (Fig. 3e) using the angle-averaged density matrix. Our results represent the first experimental verification of the mutual entanglement shared by virtually all atoms in an ensemble that contains more than a few particles.

The above results demonstrate that even with limited resources, i.e. weak atom-photon coupling, heralding schemes can be used to boost the effective interaction strength by a large factor, enabling the production of highly entangled states\cite{Agarwal2005,McConnell2013}. Furthermore, by repeated trials and feedback the entanglement generation can be made quasi-deterministic\cite{DLCZ2001,Matsukevich06}. Our approach is related to other heralded schemes for quantum communication\cite{DLCZ2001, Kuzmich03,Matsukevich06,Simon07} and entangled-state preparation\cite{Choi10,Christensen2012, ChristensenInterfere2014}, and it would be interesting to generalize the present analysis to infer characteristics of the atomic state from the measured optical signals in those experiments. We note that the same first Dicke state was created in an ensemble of up to 41 atoms with a scheme that uses many heralding photons in a strongly coupled atom-cavity system\cite{Haas2014}. In our system, the maximum atom number of $\sim 3000$ is set by the accuracy of the spin rotation, and can be increased by two orders of magnitude by better magnetic-field control\cite{Bohnet14}. The state purity $\rho_{11}$ can probably be further improved by reducing the heralding probability, and a value of $\rho_{11} > 0.73$ would be required for the Fisher information\cite{Strobel2014} to exceed that of the CSS, and enable metrological gain of up to 3~dB. The detection of two or more photons prepares Schr\"odinger cat states\cite{McConnell2013} of the atomic ensemble with more metrological gain. We expect that heralded methods can generate a variety of nearly pure, complex, strongly entangled states that are not accessible by any other means at the present state of quantum technology.

\begin{acknowledgments}
We thank M. H. Schleier-Smith, E. S. Polzik and S. L. Christensen for discussions. This work was supported by the NSF, DARPA (QUASAR), and a MURI grant through AFOSR. S.\,\'{C}. acknowledges support from the Ministry of Education, Science and Technological Development of the Republic of Serbia, through Grant No. III45016 and OI171038.
\end{acknowledgments}

R.M and H.Z. contributed equally to this work.

\end{bibunit}

\begin{bibunit}

\include{Methods_arxiv}

\end{bibunit}

\end{document}

%% file: Methods_arxiv.tex
\newcounter{sec}
\newcounter{subsec}[sec]
\setcounter{figure}{0}
\setcounter{page}{1}

\renewcommand{\figurename}{Extended Data Figure}
\renewcommand{\tablename}{Extended Data Table}

\title{Methods: Entanglement with Negative Wigner Function of Three \\Thousand Atoms Heralded by One Photon}

\author{Robert McConnell}
\affiliation{
Department of Physics, MIT-Harvard Center for Ultracold Atoms,
and Research Laboratory of Electronics, Massachusetts Institute of Technology,
Cambridge, Massachusetts 02139, USA}

\author{Hao Zhang}
\affiliation{
Department of Physics, MIT-Harvard Center for Ultracold Atoms,
and Research Laboratory of Electronics, Massachusetts Institute of Technology,
Cambridge, Massachusetts 02139, USA}

\author{Jiazhong Hu}
\affiliation{
Department of Physics, MIT-Harvard Center for Ultracold Atoms,
and Research Laboratory of Electronics, Massachusetts Institute of Technology,
Cambridge, Massachusetts 02139, USA}

\author{Senka \'{C}uk}
\affiliation{
Department of Physics, MIT-Harvard Center for Ultracold Atoms,
and Research Laboratory of Electronics, Massachusetts Institute of Technology,
Cambridge, Massachusetts 02139, USA}
\affiliation{
Institute of Physics, University of Belgrade, Pregrevica 118, 11080 Belgrade, Serbia}

\author{Vladan Vuleti\'{c}}
\affiliation{
Department of Physics, MIT-Harvard Center for Ultracold Atoms,
and Research Laboratory of Electronics, Massachusetts Institute of Technology,
Cambridge, Massachusetts 02139, USA}

\date{\today}

\maketitle

\section{Photon polarization rotation by atomic spin}
Probe laser light red-detuned by $\Delta_0/(2\pi) =-200$ MHz from the $^{87}$Rb transition $5^2S_{1/2}, F = 1$ to $5^2P_{3/2}, F' = 0$ is sent through an optical cavity containing the atomic ensemble. We first consider the case where all the atoms are coupled with equal strength to the probe light. For detuning $\Delta$ much larger than the excited state linewidth $\Gamma/(2\pi) =6.1$ MHz, the excited state manifold can be adiabatically eliminated. The vector component of the ac Stark shift is described by the Hamiltonian
\begin{equation}
\frac{H}{\hbar}=\frac{g^2}{\Delta}J_z S_z,
\label{eqn:Hvec}
\end{equation}
where $J_z = \frac{1}{2}(a^\dagger_+ a_+ -a^\dagger_- a_-) $, with $a_\pm$ the annihilation operators for photons with $\sigma^\pm$ circular polarizations. Here $2g$ is the effective single-photon Rabi frequency taking into account the multiple transitions from $5^2S_{1/2}, F = 1$ to $5^2P_{3/2}, F' = 0, 1, 2$, given by
\begin{equation}
g^2=(g_{\mathrm{1,1}}^{\mathrm{0,0}})^2 + (g_{\mathrm{1,1}}^{\mathrm{1,0}})^2 + (g_{\mathrm{1,1}}^{\mathrm{2,0}})^2-(g_{\mathrm{1,1}}^{\mathrm{2,2}})^2,
\end{equation}
where $2g_{F,m}^{F',m'}$ is the single-photon Rabi frequency between the ground state $|F = 1, m\rangle$ and the excited state $|F', m'\rangle$. As $\Delta_0$ is comparable to the hyperfine splittings of the $5^2P_{3/2}$ excited states, the interaction strength $g^2/\Delta$ is given by
\begin{equation}\nonumber
\frac{g^2}{\Delta}=\frac{(g_{\mathrm{1,1}}^{\mathrm{0,0}})^2}{\Delta_0}
+\frac{(g_{\mathrm{1,1}}^{\mathrm{1,0}})^2}{\Delta_0- \Delta_{\mathrm{1}}}
+\frac{(g_{\mathrm{1,1}}^{\mathrm{2,0}})^2}{\Delta_0- \Delta_{\mathrm{1}}-\Delta_{\mathrm{2}}}\\
-\frac{(g_{\mathrm{1,1}}^{\mathrm{2,2}})^2}{\Delta_0- \Delta_{\mathrm{1}}-\Delta_{\mathrm{2}}},
\label{eqn:gv}
\end{equation}
where $\Delta_{\mathrm{1}}/(2\pi )=72$ MHz is the hyperfine splitting between the $F' =0$ and $F' =1$ manifolds, $\Delta_{\mathrm{2}}/(2\pi)  = 157$ MHz between $F' =1$ and $F' =2$, and $\Delta/(2\pi ) =- 150$ MHz is the effective detuning when $\Delta_0/(2\pi) =-200$ MHz. The value $g^2/\Delta$ for our experiment is $2 \pi \times 0.7$ kHz.

This vector shift (\ref{eqn:Hvec}) gives rise to a $J_z$-dependent Larmor precession of the atomic collective spin $\vec{S}$ in the $xy$ plane. Consider one \ket{\sigma^\pm} photon passing through the optical cavity and causing the atomic spin to precess by phase $\pm \phi$. The characteristic atom-photon interaction time is $2/\kappa$, where $\kappa$ is the cavity linewidth, therefore the atomic phase is given by\cite{McConnell2013,TanjiAdvancesCavity2011} $\phi= g^2/(\Delta \kappa) = \eta_{\mathrm{v}} \Gamma/(4\Delta)$, where the cavity cooperativity $\eta_{\mathrm{v}}=4g^2/(\kappa \Gamma)=0.07 $. Another way to think of the Hamiltonian (\ref{eqn:Hvec}) is that the atomic spin component $S_z$ causes different phase shifts on the photon $\sigma^+$ and $\sigma^-$ components, resulting in a rotation of the linear polarization of the light. The polarization rotation angle $\vartheta=(g^2/\Delta)(S_z/2) (2/\kappa)=\phi S_z$.

In general, the incident light can introduce Raman transitions between different magnetic levels in the $F =1$ ground state manifold. We apply a bias magnetic field of 4.7 G along the cavity axis to introduce a Zeeman shift between the magnetic levels, so that the Raman coupling is off-resonant. The Larmor frequency is $\omega_\mathrm{L}/(2\pi)=3.3$ MHz, larger than the cavity linewidth $\kappa/(2\pi)= 1.0$ MHz, so that the Raman coupling can be neglected. There is also an unimportant scalar light shift, as well as a tensor light shift that gives rise to squeezing that is negligible for our experimental conditions.

\section{Experimental details}
We load an ensemble of $^{87}$Rb atoms, cooled to $T = 50 \, \mu$K, into a medium-finesse optical cavity (cavity finesse $\mathcal{F} = 5600$, linewidth $\kappa / (2 \pi) = 1.0$ MHz, cooperativity $\eta_0=0.2$ at an antinode on a transition with unity oscillator strength). The atoms are confined on the cavity axis by a far-detuned optical dipole trap at 852~nm with trap depth $U/h=20$~MHz. Characteristics of the optical cavity at the 780 nm probe laser wavelength and the 852 nm trap laser wavelength are summarized in Extended Data Table 1. One Glan-Taylor polarizing beamsplitter (Thorlabs GT5) purifies the polarization of probe light entering the cavity, while a second polarizing beamsplitter after the cavity allows us to measure the rotation of the probe light due to the atomic projection noise. Two Single Photon Counting Modules (SPCMs, models SPCM-AQRH-14-FC and SPCM-AQR-12-FC) are placed at the transmitting and reflecting ports of the polarizing beamsplitter to detect the photons. Due to the fiber coupling and finite SPCM detection efficiency at 780 nm, the overall quantum efficiency of the detection process is $q=0.3$.

\begin{table}[h]
\centering
\includegraphics[width=.5 \textwidth]{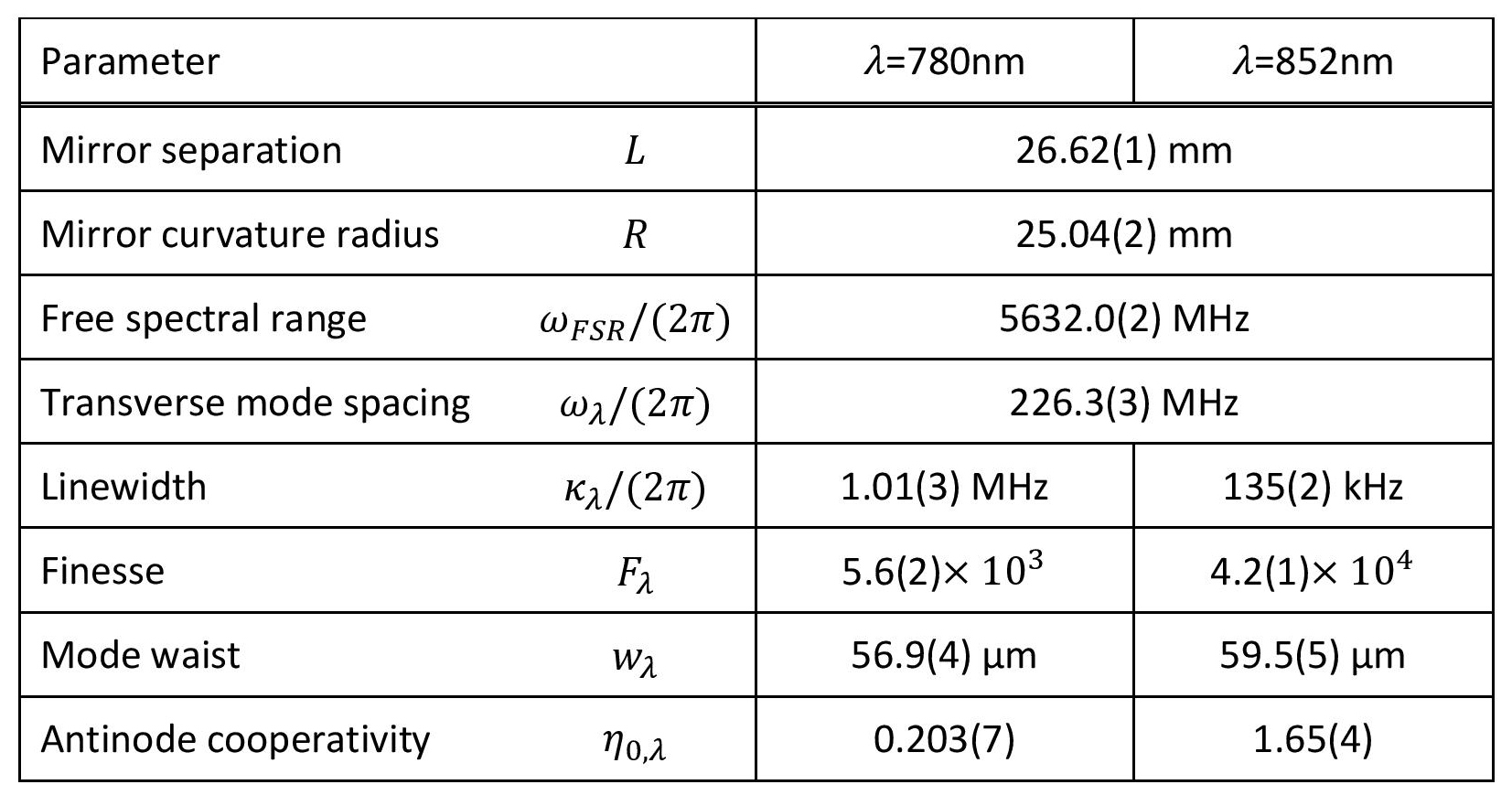}
\caption{Resonator parameters. The mode waists are calculated at the position of the atoms. Outside this table, all resonator values refer to the probe wavelength $\lambda=780$ nm.}
\label{fig:supFig1}
\end{table}

\section{Definition of effective atom number}
Atoms are optically confined at the antinodes of the 852 nm trap laser standing wave. The 780 nm probe light in the cavity forms a standing wave that is incommensurate with the trap standing wave. Consequently, the atoms experience spatially varying couplings to the probe light and rotate the probe photon polarization by different amounts. For an atom at position $z$ on the cavity axis, the cooperativity is $\eta(z)=\eta_{\mathrm{v}} \sin^2(kz)$. When $N_a$ atoms are prepared in a CSS, the atomic projection noise gives rise to fluctuations of the photon polarization rotation. The measured variance of the polarization rotation is proportional to $\frac{N_{a}}{2} \langle \eta^2(z) \rangle$ where averaging is performed over the position $z$. This variance differs by a factor of order unity from that of a CSS consisting of $N_a$ atoms uniformly coupled to the light. As described in a previous paper\cite{Schleier-Smith10}, we introduce the effective atom number $N$ and the effective cavity cooperativity $\eta$ to satisfy two conditions: that the experimentally measured variance equals that of $N$ uniformly coupled atoms, $\frac{N_{a}}{2} \langle \eta_{\mathrm{z}}^2 \rangle = \frac{N}{2} \eta^2$, and that the total amount of interaction between the atomic ensemble and the probe light is the same, i.e., $N_a \langle \eta_{\mathrm{z}} \rangle = N \eta$. To satisfy these two conditions we define the effective atom number $N = \frac{2}{3} N_a$ and the effective cavity cooperativity $\eta = \frac{3}{4} \eta_{\mathrm{v}}$. This re-scaling allows direct comparison with the well-known expressions for the uniformly coupled CSS.

As in the main paper and the rest of Methods, $S_z$ refers to the collective spin of an ensemble containing $N$ effective atoms, and therefore the atomic spin precession phase for each transmitting cavity photon is given by $\phi= \eta \Gamma/(4\Delta)=(3/4)\eta_{\mathrm{v}} \Gamma/(4\Delta)$. Note that this value $\eta =0.05 <1$ corresponds to the weak atom-cavity coupling regime. For our parameters, $\phi = 5 \times 10^{-4} \ll \phi_{CSS}=1.5 \times 10^{-2}$ where $\phi_{CSS}=\sqrt{1/(2S)}$ is the angular rms width of the CSS.

\section{Choice of the heralding photon number}
The heralding light must be weak enough that it does not introduce substantial decoherence of the desired atomic state. The fundamental shot noise between the $\sigma^+$ and $\sigma^-$ circular polarization components of the heralding light gives rise to phase broadening of the atomic state, which limits the purity of the heralded entangled state. To measure the phase broadening, heralding light pulses with variable photon number are sent into the cavity, and the variance $\Delta S_y^2$ is measured by applying a radiofrequency $\pi/2$ pulse to rotate the atomic state about the $\hat x$ direction before measuring $\Delta S_z^2$. Extended Data Fig. 1 shows the measured atomic state variance $\Delta S_y^2$ as a function of the photon number in the heralding light, in agreement with the predicted linear dependence. The heralding photon number is thus chosen to be $\sim 210$, with corresponding herald detection probability $q p= 1.5 \%$, to give fairly small phase broadening. Lower heralding photon number results in a purer heralded state, but at the expense of a lower heralding and state generation probability.

\begin{figure}[h]
\centering
\includegraphics[width=.45 \textwidth]{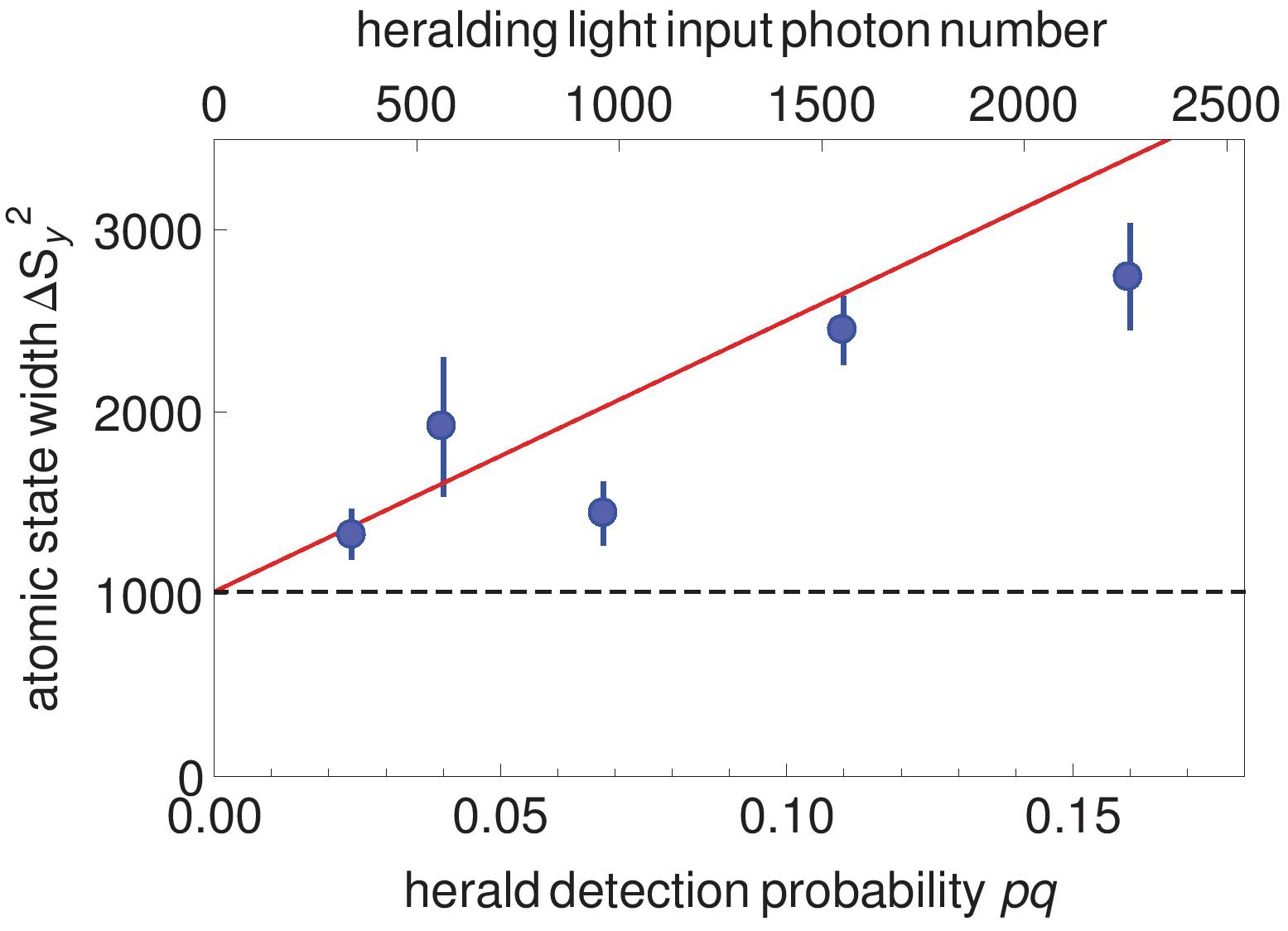}
\caption{The measured atomic state variance $\Delta S_y^2$ as a function of the heralding light photon number and corresponding probability $q p $ of detecting one photon. The solid red line is the prediction for $\Delta S_y^2$ broadened by the photon shot noise of the heralding light. The dashed black line shows the CSS variance for 2030 $F= 1$ effective atoms used in this measurement.}
\label{fig:supFig1}
\end{figure}

\section{Relation between the spin distribution $f(S_\beta)$ and the measured photon distribution $g(n_\beta)$}
To measure the atomic state spin distribution, measurement light with the same polarization $\vv$ as the heralding light is sent through the atoms, and the number of photons with the orthogonal polarization $\hh$ is measured. The measurement light contains a large number of input photons $n_\mathrm{in}=1.7 \times 10^4$ to perform destructive measurements with good signal-to-noise ratio. The photon polarization is rotated by a small angle $\vartheta = \phi S_z$ and the probability for each photon to emerge in $\hh$ is $\vartheta^2$. For a given number of input photons $n_\mathrm{in}$, the average number of detected photons with $\hh$ polarization is $\langle n \rangle= q n_\mathrm{in} (\phi S_z)^2$, where $q$ is the overall quantum efficiency. Therefore, a spin distribution $f(S_z)$ is mapped to a measured photon distribution $g(n)$. For a given $S_z$, the detected photons follow a Poisson distribution with the mean number $\langle n \rangle$, and the probability to measure exactly $n$ photons is given by
\begin{equation}
P(n, S_z)=\exp[-q n_\mathrm{in}( \phi S_z)^2]{[q n_\mathrm{in}( \phi S_z)^2]^n\over n!}.
\end{equation}
For an atomic state with the spin distribution $f(S_z)$, the photon distribution $g(n)$ is given by
\begin{eqnarray}
g(n)&=&\sum_{S_z} f(S_z) P(n, S_z) \nonumber \\
&=&\sum_{S_z} f(S_z)\exp[-q n_\mathrm{in}( \phi S_z)^2]{[q n_\mathrm{in}( \phi S_z)^2]^n\over n!}.
 \nonumber \\
\label{eqn:photonprob}
\end{eqnarray}
In order to measure the spin along a general direction, the atomic spin is rotated by an angle $\beta$ with a radiofrequency pulse prior to detection. Replacing $S_z$ by $S_\beta$ in equation (\ref{eqn:photonprob}) we write the relation between the spin distribution $f(S_\beta)$ and the measured photon distribution $g(n_\beta)$ as
\begin{eqnarray}
g(n_\beta)&=&\sum_{S_\beta} f(S_\beta) P(n_\beta, S_\beta)\nonumber \\
&=&\sum_{S_\beta} f(S_\beta)\exp[-q n_\mathrm{in}( \phi S_\beta)^2]{[q n_\mathrm{in}( \phi S_\beta)^2]^{n_\beta}\over n_\beta!}.\nonumber \\
\label{eqn:photonprob2}
\end{eqnarray}

\section{Choice of the measurement photon number}

The measurement photon number is chosen to optimize the readout quality. Extended Data Fig. 2 illustrates the dependence of readout on the input measurement photon number $n_\mathrm{in}$ by showing how the reconstructed distributions $f(S_z)$ change as $n_\mathrm{in}$ is varied (the method of reconstruction is discussed later). When the photon number is small, there is large detection noise due to photon shot noise, reflected as the large error band. With increasing photon number, the photon scattering by atoms into free space increases and the atomic state is more strongly perturbed, therefore the ``dip'' at $S_z=0$ becomes less distinct. To balance these two competing effects, the optimized atomic-state-measurement photon number is set to $1.7 \times 10^4$.

\begin{figure*}[ht]
\centering
\includegraphics[width=.75 \textwidth]{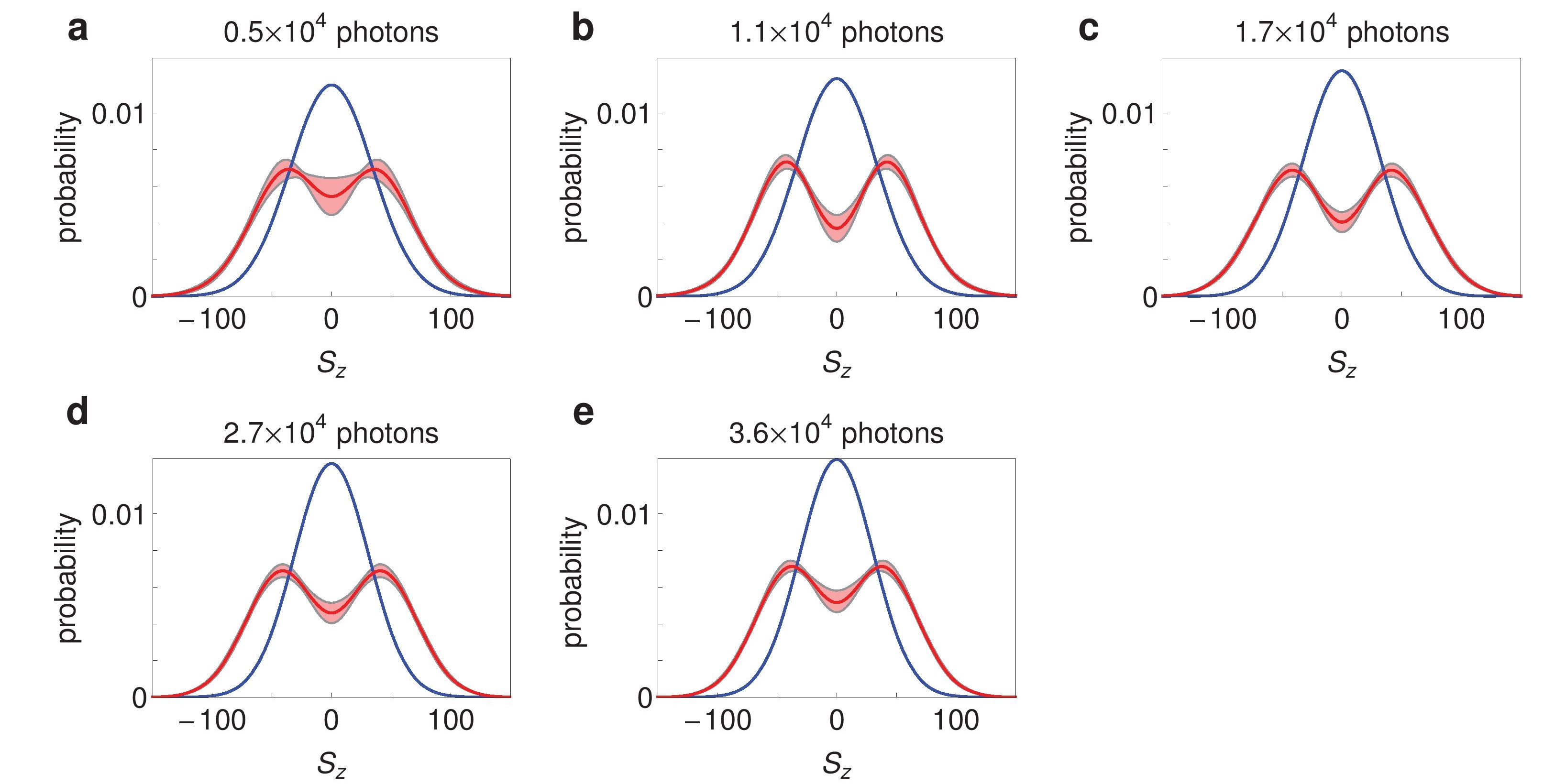}
\caption{Dependence of the reconstructed distribution of collective spin $S_z$ on the measurement photon number, as illustrated by reconstructed spin distributions for photon numbers (a) $0.5 \times 10^4$, (b) $1.1 \times 10^4$, (c) $1.7 \times 10^4$, (d) $2.7 \times 10^4$, (e) $3.6 \times 10^4$. Blue lines correspond to the CSS and red lines correspond to the heralded states. The shaded area indicates an uncertainty of 1 standard deviation.}
\label{fig:supFig2}
\end{figure*}

\section{Subtracting background photon counts}

Due to the residual polarization impurity of the measurement light, there are a small number of background photon counts even when there are no atoms. The background counts account for about 4\% of the photon signal of the heralded state. We independently measure the background photon distribution and subtract it from the directly measured atomic signal to obtain $g(n_\beta)$. If we were not to correct for these background counts, we would overestimate the density matrix population $\rho_{11}$ by 10\%.

\section{Reconstruction of the density matrix}

Using the measured photon distributions $g(n_\beta)$ for all four angles $\beta = 0, \pi/4, \pi/2, 3\pi/4$, the density matrix $\rho$ of the heralded state can be reconstructed.

As the entangled state maintains $0.99^{+0.01}_{-0.02}$ contrast, the length of the total spin $S\approx N$ and we can express the density matrix in the basis of Dicke states $|m\rangle_x$ along the $\hat x$ direction
\begin{eqnarray}
\rho&=&\rho_{00}|0\rangle_x\langle0|_x+\rho_{11}|1\rangle_x\langle1|_x
+\rho_{01}|0\rangle_x\langle1|_x+\rho_{10}|1\rangle_x\langle0|_x  \nonumber \\
&&+\rho_{22}|2\rangle_x\langle2|_x+\rho_{02}|0\rangle_x\langle2|_x
+\rho_{20}|2\rangle_x\langle0|_x+\ldots.
\end{eqnarray}
The spin distribution $f (S_\beta)$ can be written as a function of atom number $N$ and the density matrix elements $\rho_{00}, \rho_{11}$, etc:
\begin{eqnarray}
&&f(S_\beta, \rho, N)=\langle S_\beta|\rho|S_\beta\rangle \nonumber \\
&=&\rho_{00}G(0,S_\beta)G^*(0,S_\beta)+\rho_{11}G(1,S_\beta)G^*(1,S_\beta) \nonumber \\
&&+\rho_{01}G(0,S_\beta)G^*(1,S_\beta) +\rho_{10}G(1,S_\beta)G^*(0,S_\beta)\nonumber \\
&&+\rho_{22}G(2,S_\beta)G^*(2,S_\beta) +\rho_{02}G(0,S_\beta)G^*(2,S_\beta) \nonumber \\
&&+\rho_{20}G(2,S_\beta)G^*(0,S_\beta)+\ldots.
\label{eqn:frho}
\end{eqnarray}
Here $G(m,S_\beta)=\langle S_\beta|m\rangle_x$ is the wavefunction of Dicke state $|m\rangle_x$ in the representation of spin component $S_\beta$ and is given by
\begin{eqnarray}
G(m,S_\beta,N)&=&{1\over\sqrt{2^m m!}}\left({1\over\pi N}\right)^{1/4}\times \nonumber \\
&&\times e^{i m \beta-S^2_\beta/(2N)}H_m\left(\sqrt{1\over N}S_\beta\right),
\end{eqnarray}
where $H_m(x)$ is the $m$th order Hermite polynomial and $N$ is the atom number.
Using equation (\ref{eqn:photonprob2}), we write the theoretically predicted photon distribution $g_{th}(n_\beta)$ as a function of the density matrix $\rho$, atom number $N$ and input photon number $n_\mathrm{in}$
\begin{eqnarray}
&&g_{th}(n_\beta,\rho, N, n_\mathrm{in})=\sum_{S_\beta} f_{th}(S_\beta,\rho, N) P(n_\beta, S_\beta)\nonumber \\
&=&\sum_{S_\beta} f_{th}(S_\beta,\rho, N)\exp[-q n_{in}(S_\beta \phi)^2]{[q n_{in}(S_\beta \phi)^2]^{n_\beta}\over n_{\beta}!}.\nonumber \\
\label{eqn:countsf}
\end{eqnarray}
We independently measure the input photon number $n_\mathrm{in}$ and find the atom number $N$ by fitting the photon distributions of the CSS, whose only non-zero density matrix element is $\rho_{00}=1$. The fitted atom numbers $N$ for different angles $\beta$ agree within 15\% with the values independently measured from the shift of the cavity resonance. We then use the density matrix $\rho$ of the heralded state as the only free parameter, to fit the theoretical distributions $g_{th}(n_\beta)$ to the measured photon distributions $g(n_\beta)$ along all four angles $\hat{\beta}$. We do this by minimizing the least squares deviation $D$ weighted by the error $\sigma_{g}$ of $g(n_\beta)$, given by
\begin{equation}
D=\sum_{\beta}\sum_{n\geq 0}\left[\frac{g_{th}(n_\beta,\rho)-g(n_\beta,\rho)}{\sigma_{g}}\right]^2.
\end{equation}

Since the photon distributions $g(n_\beta)$ measure $S_\beta^2$, we can obtain the even terms of the density matrix ($\rho_{mn}$ where $m+n$ is even) and are not sensitive to the odd terms. Because the overall heralding probability is $pq=1.5\%$, the higher-order Dicke state components are exponentially suppressed. We fit the density matrix up to Dicke state $|4\rangle_x$. The fitted values $\rho_{22}=0.03 \pm 0.02, \rho_{33}=0.02 \pm 0.01, \rho_{44}=0.01 \pm 0.01$ agree with the theoretical expectation\cite{McConnell2013} for our system.

From the fitted density matrix $\rho$ (with coherence terms) we obtain the spin distributions $f(S_\beta)$ using (\ref{eqn:frho}) for different angles $\beta$, as shown in Fig. 2e-h of the main text.

To reconstruct the Wigner function for the spin state on the Bloch sphere\cite{McConnell2013,Dowling1994}, we convert $\rho$ from the Dicke state basis into the spherical harmonic basis and obtain the normalized Wigner function according to
\begin{equation}
W(\theta,\phi)=\frac{1}{\sqrt{2 S/\pi}}\sum^N_{k=0}\sum^k_{q=-k}\rho_{kq}Y_{kq}(\theta,\phi),
\end{equation}
where the terms $\rho_{kq}$ represent the density elements in the spherical harmonic basis and $Y_{kq}(\theta,\phi)$ are the spherical harmonics, with $\theta,\phi$ being the polar and azimuthal angles on the Bloch sphere respectively. The normalization factor $\sqrt{2 S/\pi}$ is chosen such that the CSS has $W(\frac{\pi}{2},0) = 1$. Note that, in the limit of large atom number, this normalization also means that the pure first excited Dicke state has $W(\frac{\pi}{2},0) = -1$, and generally the value of the Wigner function on the $\hat{x}$ axis depends only on the populations $\rho_{nn}$ such that $W(\theta=\frac{\pi}{2},\phi=0)= \sum_{n} (-1)^{n} \rho_{nn}$.

\section{Measurement of mean value of $S_z$}
The measured photon distributions $g(n_\beta)$ do not give information about the density matrix odd terms ($\rho_{mn}$ where $m+n$ is odd). In order to bound the odd terms we verify that the heralding process does not displace the produced heralded state relative to the CSS. This is accomplished by performing a measurement with a probe beam polarized at 45 degrees relative to $\vv$, such that the difference between the measured $\hh$ and $\vv$ photon numbers is proportional to $S_z$. We find a heralding-light-induced shift $\delta \aver{S_z} = -0.2 \pm 1.6$, consistent with zero, and very small compared to the CSS rms width $(\Delta S_{z})_{\mathrm{CSS}} \approx 30$. Therefore we set the odd terms of the density matrix to zero in Fig. 3b-d.

\section{Entanglement depth for finite contrast}
Entanglement depth is defined as the minimum number of entangled particles in an ensemble. A fully separable pure state can be written as $\ket{\varphi}= \ket{\varphi_1}\otimes ... \otimes \ket{\varphi_N}$, where $N$ is the atom number. A pure $k$-producible state can be written as $\ket{\varphi}= \ket{\varphi_1^{1,...,k_1}}\otimes ... \otimes \ket{\varphi_M^{1,...,k_M}}$, where $k_1,...,k_M \leq k$, $k_1+...+k_M = N$. If a state cannot be written as a pure $(k-1)$-producible state or a mixed state of $(k-1)$-producible states, then it has entanglement depth of at least $k$.

We slightly generalize the entanglement criterion derived in Ref.\cite{Haas2014} to take into account the finite contrast $\mathcal{C}$ of the collective atomic spin in our experiment. The derivation in Ref.\cite{Haas2014} considers the case in the fully symmetric Dicke subspace of $N$ atoms, and finds that for a $k$-producible state the maximum population of the first Dicke state $\rho_{11}$ ($P_1$) as a function of the CSS population $\rho_{00}$ ($P_0$) is
\begin{eqnarray}
\max_{P_0}P_1&=&{P_0\over N}\max\left[\sqrt{k}\max_{\prod_{i=1}^{M-1}a_i=x}F_{M-1}(a_1,\ldots, a_{M-1})\right.  \nonumber \\
&+&\left.\sqrt{k'}F_1(\sqrt{P_0}/x)\right]^2.
\label{eqn:ksep}
\end{eqnarray}
Here $M= [N/k]$, $k'= N- k(M-1)$, and $F_n(a_1,\ldots,a_n)=\sum^n_{i=1}{\sqrt{1-a^2_i}\over a_i}$. Equation (\ref{eqn:ksep}) is generally not a concave function of $P_0$. In order to obtain the upper bound for mixed states, denote the concave hull of the right side of equation (\ref{eqn:ksep}) as $\mathcal{B}(P_0,k,N)$. We define $\mathcal{B}(P_0,k,N) = B(P_0,k,N)/N$. Note that when $N_1<N_2$, $B(P_0,k,N_1)\leq B(P_0,k,N_2)$.

The heralded state we produce does not necessarily retain perfect contrast, so the state can be a mixture of different total spins $S=N, N-1, ..., N(1-\epsilon)$, with $\epsilon\sim1\%$. The contrast loss is mainly caused by the decoherence between $F=1$ magnetic sublevels, and the free space scattering of the heralding light by the atoms. We decompose the density matrix $\rho$ into the total spin basis
\begin{equation}
\rho=\sum_{i=0}^{\epsilon N} w_i \rho_{N-i}.
\end{equation}
Here $\rho_{N-i}$ is the density matrix in the subspace of total spin $S = N-i$, $w_i$ is the weight for each $\rho_{N-i}$ and $\sum w_i=1$. For each $\rho_{N-i}$,
\begin{eqnarray}
\mathcal{B}(P_0,k,N-i)&=&B(P_{0,N-i},k,N-i)/(N-i) \nonumber \\
&\leq& B(P_{0,N-i},k,N)/(N-i).
\label{eqn:inequ1}
\end{eqnarray}
Here $P_{0,N-i}$ is the probability for the state to be found in the ground state in the subspace of total spin $N-i$.

Measurements of the spin distributions do not allow us to determine the total spin of the system at single-atom resolution. We define populations of the CSS and the first Dicke state by
\begin{eqnarray}
P_0&=&\sum_{i=0}^{\epsilon N} w_i P_{0,N-i}, \\
P_1&=&\sum_{i=0}^{\epsilon N} w_i P_{1,N-i}.
\end{eqnarray}
The upper bound of $P_1$ is given by
\begin{eqnarray}
\max_{P_0}P_{1}&\leq&\sum_{i=0}^{\epsilon N} w_i \max_{P_{0,N-i}}P_{1,N-i} \nonumber \\
&\leq& \sum_{i=0}^{\epsilon N} w_i B(P_{0,N-i},k,N-i)/(N-i).
\end{eqnarray}
Using equation (\ref{eqn:inequ1}) and the fact that $B(P_0,k,N)$ is a concave function of $P_0$ we have
\begin{eqnarray}
\max_{P_0}P_{1}&\leq&\sum_{i=0}^{\epsilon N} w_i B(P_{0,N-i},k,N)/(N-\epsilon N)\nonumber \\
&\leq&{1\over (1-\epsilon)N}B\Big(\sum_{i=0}^{\epsilon N}w_i P_{0,N-i},k,N\Big)\nonumber \\
&=&{1\over \mathcal{C} }\mathcal{B}(P_0,k,N).
\end{eqnarray}
Here $\mathcal{C}$ is the contrast of the collective spin. Comparing to Ref.\cite{Haas2014}, the result is modified by a factor $1/\mathcal{C}$. In our experiment, $\mathcal{C}=0.99^{+0.01}_{-0.02}$, so the effects of finite contrast on entanglement depth are minimal.

\section{Entanglement depth in terms of the actual atom number}
In the experiment the atoms have spatially varying coupling to the probe light. However, the criterion in Ref.\cite{Haas2014} is derived for the case where atoms are equally coupled to the light. Here we generalize the entanglement criterion to our experimental conditions and prove that the sample-averaged fractional entanglement depth for the ensemble containing 3100 actual non-uniformly coupled atoms is the same as that of 2100 uniformly coupled effective atoms. Consider an ensemble of $N_a$ actual atoms where each atom $j$ has spin component $f_{z,j}$ and cooperativity $\eta_j$. The effective total spin of the ensemble is $S_z$ and the effective cooperativity is $\eta$, so that
\begin{equation}
S_{z}\eta=\sum_{j=1}^{N_a} f_{z,j}\times\eta_j .
\end{equation}

As mentioned in the main paper, the ideal heralded state $|\psi_1\rangle$ (the first Dicke state of non-uniformly coupled atoms) is the destructive interference of two slightly displaced CSSs $\ket{\pm\phi}$ and can be written as
\begin{eqnarray}
|\psi_1\rangle &\propto& \ket{\phi}-\ket{-\phi}\nonumber\\
&=&\left[e^{iS_{z}\eta \Gamma/(4 \Delta)}-e^{-iS_{z}\eta\Gamma/(4 \Delta)}\right]|\psi_0\rangle \nonumber\\
&=&\left[e^{i \Gamma/(4 \Delta)\sum_{j=1}^{N_a}  f_{z,j}\eta_j}-e^{-i \Gamma/(4 \Delta)\sum_{j=1}^{N_a} f_{z,j}\eta_j}\right]|\psi_0\rangle,\nonumber \\
\end{eqnarray}
where $|\psi_0\rangle$ is the initial CSS along $\hat{x}$.
By expanding the exponent to first order and using $f_z=(f_{+,x}-f_{-,x})/(2i)$, we get
\begin{equation}
|\psi_1\rangle=\left(\sum_{j=1}^{N_a} \eta^2_j\right)^{-1/2}\sum_{j=1}^{N_a} \eta_j\left[\prod_{j'\neq j}|0_{j'}\rangle_x\right]\otimes|1_j\rangle_x,
\label{Dicke-nonuni}
\end{equation}
where $|0_j\rangle_x$ and $|1_j\rangle_x$ are the single-particle spin eigenstates along $\hat{x}$ of the atom $j$. For a fully separable state $|\varphi \rangle =\prod_{j=1}^{N_a}\left(\alpha_j|0_j\rangle_x+\beta_j|1_j\rangle_x+\ldots\right)$ the population $P_1=|\langle \varphi| \psi_1 \rangle|^2$ is given by
\begin{equation}
P_1=\left(\sum_{j=1}^{N_a} \eta^2_j\right)^{-1}\left|\sum_{j=1}^{N_a} \eta_j\beta_j\prod_{j'\neq j}\alpha_{j'}\right|^2.
\end{equation}
The expression for $P_1$ is similar to that in Ref.\cite{Haas2014} and differs by the additional weight factor $\eta_j$. When the real atom number $N_a \gg 1$, the upper bound of $P_1$ for the fully separable state $|\varphi \rangle$, $\mathcal{B}(P_0, N_a)$, as a function of the population $P_0 = |\langle \varphi| \psi_0 \rangle|^2$, is the same as Ref.\cite{Haas2014}, and independent of $N_a$.

Next consider a state which can be factorized into two subsets $\ket{\varphi}= \ket{\varphi_1^{1,...,k_1}}\otimes \ket{\varphi_M^{1,...,k_2}}$ where $k_1 + k_2 = N_a$. Each $\ket{\varphi_{i=1,2}}$ can be expanded as
\begin{equation}
\ket{\varphi_{i}} = a_i \ket{\psi_0^{k_i}} + b_i \ket{\psi_1^{k_i}}+...,
\end{equation}
where $\ket{\psi_0^{k_i}}$ is the CSS containing $k_i$ atoms, and $\ket{\psi_1^{k_i}}$ is given by \ref{Dicke-nonuni} with $N_a$ replaced by $k_i$.
The populations $P_0 = |\langle \varphi| \psi_0 \rangle|^2$ and $P_1 =|\langle \varphi| \psi_1 \rangle|^2$ are given by
\begin{eqnarray}
P_0&=& |a_1|^2 |a_2|^2, \nonumber\\
P_1&=&\left(\sum_{j=1}^{N_a} \eta^2_j\right)^{-1}\left|a_2 b_1  \sqrt{\sum_{j=1}^{k_1}\eta_j^2} + a_1 b_2 \sqrt{\sum_{j=k_1+1}^{N_a} \eta_j^2}\right|^2 \nonumber \\
\label{P-nonuni}
\end{eqnarray}

The expression for $P_1$ recovers that of Ref.\cite{Haas2014} when $\eta_j =1 $. When $k_1, k_2$ and $ N_a$ are large, we take the ensemble averages $\sum_{j=1}^{k_1}\eta_j^2 =k_1 \langle \eta^2 \rangle$, $ \sum_{j=k_1+1}^{N_a} \eta_j^2 = k_2 \langle \eta^2 \rangle$ and $\sum_{j=1}^{N_a}\eta_j^2 =N_a \langle \eta^2 \rangle$. Therefore the bound of $P_1$ in equation (\ref{P-nonuni}), $\mathcal{B}(P_0, k_a=\max\{k_1,k_2\}, N_a)$, is the same as $\mathcal{B}(P_0, k, N)$ for uniformly coupled atoms when $k_a/N_a = k/N$. This proves that the average fractional entanglement depth for the ensemble containing 3100 actual non-uniformly coupled atoms is the same as that of 2100 uniformly coupled effective atoms, thus in our system a minimum of 1970 out of 2100 effective atoms or 2910 out of 3100 real atoms are mutually entangled.

It might seem as if the addition of $N_\mathrm{w} \gg N$ weakly coupled atoms (coupling strength $\eta_\mathrm{w}$) to the system would increase the entanglement depth without having physical consequences as long as $N_\mathrm{w} \eta_\mathrm{w}^2 \ll N \eta^2$. However in this case the uncertainty $\Delta \mathcal{N'}$ on the entanglement depth also increases, given by $\frac{\Delta \mathcal{N'}}{N_\mathrm{w}} = \frac{\Delta \mathcal{N}}{N}\frac{N \eta^2}{N_\mathrm{w} \eta_\mathrm{w}^2} \gg \frac{\Delta \mathcal{N}}{N}$, so as to be consistent with the entanglement depth $ \mathcal{N}$ prior to adding the weakly coupled atoms. Atoms that do not change the observed spin distribution have no effect on the entanglement depth.